\documentclass[epj,referee]{svjour}
\usepackage{graphics}
\usepackage{graphicx}
\usepackage{graphics}
\usepackage{amsmath}
\usepackage{amssymb}
\usepackage{epsf}
\usepackage{epstopdf}
\usepackage{color}
\usepackage{setspace}
\newcommand{\be}{\begin{equation}}
\newcommand{\ee}{\end{equation}}
\newcommand{\bea}{\begin{eqnarray}}
\newcommand{\eea}{\end{eqnarray}}
\newcommand{\bdm}{\begin{displaymath}}
\newcommand{\edm}{\end{displaymath}}
\newcommand{\bit}{\begin{itemize}}
\newcommand{\eit}{\end{itemize}}
\newcommand{\ben}{\begin{enumerate}}
\newcommand{\een}{\end{enumerate}}
\begin{document}
\title{A Lattice-Boltzmann model for suspensions of self-propelling colloidal particles}
\author{Sanoop Ramachandran \inst{1}, P. B. Sunil Kumar \inst{1,2} \and I. Pagonabarraga \inst{3}
}
\institute{Department of Physics, Indian Institute of Technology Madras, Chennai 600036, India \and MEMPHYS--Center for Biomembrane Physics, University of Southern Denmark, DK-5230, Denmark
\and Department de Fisica  Fonamental, Universitat de Barcelona,
08028-Barcelona, Spain}
\date{Received: date / Revised version: date}
\abstract{ We present a Lattice-Boltzmann method for simulating self-propelling ( active) colloidal particles
 in two-dimensions. Active particles with  symmetric and asymmetric force distribution on its surface 
  are considered. The velocity field generated by a  single active particle, changing its orientation 
  randomly,  and the different time scales involved are  characterized  in detail.  The steady state speed distribution in the fluid, resulting from the activity, is shown to deviate considerably from the equilibrium distribution.
%end of PACS codes
} %end of abstract
\authorrunning{S. Ramachandran, P.B.Sunil Kumar and I. Pagonabarraga}
\titlerunning{Lattice-Boltzmann for active colloids}
\maketitle
\section{Introduction}
\label{intro}

Motion at low Reynolds number is an intriguing phenomena. How does a  
micron sized bacteria moving in a fluid change the velocity field of  the fluid around it?
Does this local stirring change the viscocity and temperature of the fluid?
Such questions have gained importance, thanks to recent  
experiments on the motion of bacteria in two dimensional fluid films
~\cite{wu-libchaber-00,goldstein}. Theoretical work on active particle suspensions has addressed  the organization of active particles and their collective motion~\cite{viscek-95,toner-tu-98}, as well as  pattern formation induced by activity~\cite{lee-kardar-01,sumitra-sunil-gautam-04}. The peculiarities of the rheological response of active suspension has also been analyzed~\cite{madan-sriram-04,aditi-sriram-02}.

The collective, hydrodynamic behavior of active interacting organisms has been analyzed disregarding the  dynamics of the embedding solvent~\cite{toner-tu-98}. In doing so, the microscopic details of propulsion and particle interactions were hidden in  the phenomenological coefficients.  In a more microscopic description, Ramaswamy and co workers model the active particles as point force dipoles~\cite{aditi-sriram-02} and study the possibility of nematic ordering in such a system. More complicated force distributions on the active particles are difficult to carry out in theoretical treatments. In this paper we present a Lattice-Boltzmann simulation scheme  for active colloidal particles. The model can be used to study the rheological behaviour and self-organization in active particle suspensions~\cite{ramachandran-06}.

The Lattice-Boltzmann (LB)  method relies on the fact that the macroscopic
dynamics of fluid  flow does not depend on the microscopic details of the solvent. LB is a mesoscopic  kinetic model that captures the collective macroscopic dynamics of fluid flow.  The method involves the discretization of both  space and time, thereby reducing the number of degrees of freedom of the system, which in turn leads to a reduced computational effort. By introducing simplified kinetic rules one captures the essential macroscopic behaviour to a high degree of accuracy. 

Over the years, LB has been used to simulate a wide variety of systems~\cite{succi}. 
It provides a faithful discretization of Navier-Stokes equation near  incompressible-fluid flow, 
and has been used for large scale fluid dynamic simulations~\cite{lockard-02}.   It has also 
proved to be a flexible method for the study of complex fluid flows and has been applied to 
study  porous medium flows, binary fluid separation kinetics as well as colloidal and 
polymer suspensions~\cite{ignacio} .

Among the different proposals to simulate rigid particles suspended in a fluid 
(such as a colloidal suspension) in LB we will follow the method introduced by 
Ladd~\cite{ladd-94} which defines the solid through the links that join particle and 
fluid nodes, called the boundary links.  The  suspended objects interact with the 
surrounding fluid through the conditions specified at the boundary links. In the 
original version, the fluid also occupies the interior of the  closed surface defined by 
the boundary links.  The inertia of the inner fluid may affect the dynamics of suspended 
colloids at short time scales. More recently,  Nguyen and Ladd~\cite{nguyen-ladd-02},
have introduced a variant of this method which excludes the fluid at the interior of the 
colloid. While for single phase fluids such a variant is not crucial, excluding the fluid 
from inside the colloids cannot be avoided in the case of fluid mixtures 
~\cite{cates-adhikari-04},  where the concentration gradients in the neighbourhood of 
the colloid must be accounted for appropriately.  In both descriptions the appropriate 
stick boundary conditions of the fluid at the solid surface are accomplished by bouncing 
back the incoming fluid densities in the frame of reference of the moving particle.

  In this paper we present the first  two-dimensional  LB model
of a  suspension in which the solid particles are driven by internal 
forces.  In section II, we describe in detail the method used to simulate the 
active motion of the particles through appropriate multipolar force distributions. Section
III concentrates on the results for a single active particle, in the absence of thermal 
fluctuations, when the force distribution 
does not induce any net motion - a "shaker". Section IV describes the peculiarities of a 
particle that can displace - a "mover".  We conclude with section V, where we discuss the 
main results obtained and indicate future directions.

\section{Active Colloid Model}
\label{model}
 The LB approach has been widely used to study the dynamics of colloidal suspensions~\cite{ladd-94}.  We use the algorithm proposed by Ladd for a  colloidal suspension  and extend it to 
incorporate active particles.  In the following section we briefly  describe the discretized LB approach.

\subsection{Fluid Lattice Model}

 The central dynamic quantity in LB is the  one particle distribution function, $n_i ({\bf r},{\bf c}_i, t)$, 
which corresponds to the density of fluid particles at the lattice site 
${\bf r}$  moving with velocity ${\bf c}_i$.   All the length and time scales used are in terms of lattice units, which implies that the particles move a unit distance per unit time.  

  Having defined the distribution function, $n_i ({\bf r},{\bf c}_i, t)$,  one can 
recover the hydrodynamic fields as its various moments. In particular, 
the zeroth moment gives the local density of the fluid $\rho = \sum_i n_i \label{1}$, 
the local momentum of the fluid ${\bf j} = \rho {\bf u} =\sum_i n_i {\bf c}_i $. 
 while the momentum flux is given by the second moment,
$	{\bf \Pi} = \sum_i n_i {\bf c}_i {\bf c}_i  $, 
where ${\bf c}_i {\bf c}_i$ is a diadic product.

\begin{figure}[h]
\centering
\includegraphics[scale=0.5]{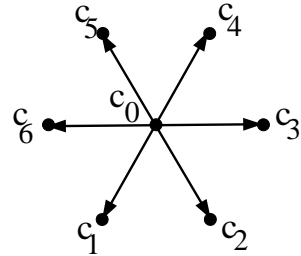}
\caption{The seven velocity directions $c_i$, where $i = 0,1...6$.}
\label{directions}
\end{figure}

  In LB, at each time step the distribution at every lattice site is 
updated in two substeps; a propagation and a collision event.  In the propagation event, 
$n_i$ is transferred to the corresponding neighbouring site along the link corresponding 
to its velocity, ${\bf c}_i$.  In the collision substep, the distribution at every site is
relaxed toward its local equilibrium distribution in a way that  ensures local mass and 
momentum conservation. The overall dynamics can be 
expressed as

\be 	
n_i({\bf r} + {\bf c}_i, t + 1) =  n_i ({\bf r}, t) + \sum_j \mathsterling_{ij}(n_j - n_j^{eq}),	
\ee

where the first term in the r.h.s. expresses the propagation, and  the second term 
corresponds to collisions.  $\mathsterling_{ij}$ is the linearized collision
matrix while   $n_j^{eq}$ is the equilibrium distribution which depends on the local values
 of $\rho$ and ${\bf u}$. For simplicity's sake, we restrict ourselves to the particular case in which $\Pi$ relaxes to its equilibrium value in one time step. In this case the equilibrium distribution function can be simplified to~\cite{ladd-01}

\be 	n_i^{eq} = \rho \left [ a_0^{c_i} + a_1^{c_i} {\bf u} \cdot {\bf c}_i \right ],\ee
 
with fluid viscosity set to $\eta=\frac{1}{8}\rho$. The coefficients in the equilibrium distribution function  are obtained
imposing isotropy, mass and momentum conservation.   The   post-collision distribution can now be expressed as 
\be
n_i+\Delta_i(n_i)=a_0^{c_i}\rho+a_1^{c_i}j_\alpha c_{i\alpha}
\ee
for zero Reynolds number flows. The results obtained in this paper correspond to a D2Q7 lattice,  which means a two dimensional triangular lattice, with seven velocity directions at every site, see figure~\ref{directions}.  For this lattice, the amplitudes of the equilibrium distribution are
\be 
a_0^0=1-2c_s^2,
a_0^1=c_s^2/3,
a_1^1=1/3,
\ee
where $c_s$ is the speed of sound in the fluid. Since the distribution function 
must be positive, $c_s$ is restricted to $0 \le c_s \le \frac{1}{\sqrt{2}}$.
For convenience, the density of particles at every point  moving in each direction is set 
to unity, leading to $\rho^{eq} =7$ at every lattice site.

\subsection{Solid particles}

\begin{figure}[h]
\centering
\includegraphics[scale=0.5]{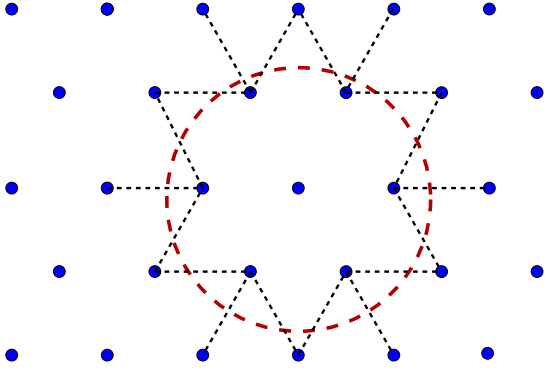}
\caption{Discrete representation of a solid particle along with the boundary links, 
when its center of mass is displaced from the lattice node.  The dashed circle represents the particle.}
\label{discrete}
\end{figure}

  We follow Ladd's approach and define the 
solid particles through the set of boundary links which define a closed surface. These boundary 
 links join neighbouring fluid and solid nodes. This procedure allows for  simulation objects of  any shape, 
as shown in figure~\ref{discrete}. As it moves, the center of mass of the particle makes a smooth trajectory in space. It is worth noting 
that the shape of the object changes as it moves, since the number of boundary links will 
in general change. Such lattice artifacts can be systematically reduced by increasing the particle's size. 

Apart from the propagation and relaxation events which define the LB dynamics, the density 
of fluid particles moving  along a boundary link are 
bounced-back to enforce  stick boundary conditions as described below.
  Lattice nodes on either side of the boundary surface are treated on the same footing,
so that the fluid fills the space inside and outside the solid particles; 
the interior fluid being kept only for computational convenience. If needed, there are alternative approaches which exclude the interior fluid~\cite{nguyen-ladd-02}.

  At each boundary link there are two distributions $n_i({\bf r},t)$ and 
$n_{i^\prime}({\bf r}+{\bf c}_i,t)$, corresponding to velocities ${\bf c}_i$ and 
${\bf c}_{i^\prime}$ (${\bf c}_{i^\prime}=-{\bf c}_i)$ parallel to the boundary link connecting
${\bf r}_i$ and ${\bf r}_i + {\bf c}_i$.  If the boundary is stationary, the incoming 
distributions are simply reflected back in the direction they come from.  If the boundary 
is moving, upon reflection, a part of the momentum is transferred across 
the boundary node. These boundary node update rules can be written as
  \be	
n_i({\bf r}+{\bf c}_i,t)=n_{i^\prime}({\bf r}+{\bf c}_i,t)+2a_1^{{\bf c}_i}
 \rho {\bf u}_b \cdot  {\bf c}_i 	
\ee
\be	
n_i({\bf r}_{i^\prime},t)=n_i({\bf r},t) - 2a_1^{{\bf c}_i} \rho {\bf u}_b \cdot  
{\bf c}_i	
\ee

where the boundary link velocity, ${\bf u}_b$, is  calculated by
\be	{\bf u}_b = {\bf U} + {\bf \Omega} \times ({\bf r} +  \frac{{\bf c}_i}{2} - {\bf R}). 	\ee
 Here $\bf U$ is the center of mass velocity of the particle, ${\bf \Omega}$
 is the angular velocity, ${\bf r} + {\bf c}_i/2\equiv {\bf r}_b$ is the position of the boundary link while 
  ${\bf R}$ is center of mass position.
  
  Due to the bounce back, the fluid exerts a net force on the particle.  As a result, there is an exchange of momentum at the boundary, but 
the total momentum of the particle and the fluid combined is conserved.  Using this 
condition, the net force and momentum acting on the particle can be calculated.  The force 
and the torque at every node can be expressed as,

\begin{eqnarray}
 {\bf F}_b({\bf r}_b)&=&2\left[ n_i({\bf r},t)\right.-\left. n_{i^\prime}({\bf r}+{\bf c}_i,t) \right. \nonumber \\
          && \left.-2 a_1  \rho({\bf r}) {\bf u}_b \cdot  {\bf c}_i \right] {\bf c}_i  \label{force}\\
 {\boldsymbol \tau}_b& =&{\bf r}_b \times {\bf F}_b ({\bf r}_b)   \label{torque} 
 \end{eqnarray}
  which  give rise to net force and torques acting on the particle,
\begin{eqnarray} 
{\bf F}({\bf r},t+1)&=&\sum_{{\bf r}_b} 2 \left [n_i({\bf r},t)  
             -  n_{i^\prime}({\bf r}+ {\bf c}_i,t)\right. \nonumber \\
&& \left. -2a_1{\bf c}_i\rho({\bf r}){\bf u}_b \cdot
 {\bf c}_i \right ] {\bf c}_i	
\label{eqn:fb}
\end{eqnarray}

\be
{\boldsymbol \tau}({\bf r}_b,t+1)=\sum_{{\bf r}_b}{\bf r}_b \times {\bf F}_b({\bf r}_b,t+1)	
\ee
where the sums run over all the boundary links.
Then the particle linear velocity and angular velocity are updated using 
\be  {\bf U}(t+1) = {\bf U}(t) + 2 M^{-1} {\bf F}(t+1) 	\ee
\be  {\bf \Omega}(t+1) = {\bf \Omega}(t) + 2 I^{-1} {\boldsymbol \tau}(t+1) \ee
where $M$ is the mass of the particle and $I$ is its moment of inertia. 

\subsection{Self-propulsion} \label{activity}
So far we have been discussing colloidal particles that are passive. In presence of 
thermal fluctuations the boundary interaction described above will ensure that 
the particles execute Brownian motion. 
We will now modify the above algorithm in order to incorporate 
the self-propulsion of the particles, their ``activity".  Since by Newton's third law, 
the net force acting on a self-propelled particle must be zero, we thus have to distribute 
forces, at the boundary nodes, such that there is no net force exerted by the particle.
 To this end, we divide the particle into two parts and  distribute equal and opposite forces in the two halves.  The  simplest form for such a self-propulsion is a force dipole~\cite{brennen-77}.  

To implement this scheme, first a random direction is chosen, denoted by the unit
vector $\hat{h}$.  The particle is then divided, perpendicular to this axis, into two 
parts at a distance $d$ from its geometric center.  We then assign  forces to each of the boundary nodes such that the net force in one sub-domain is equal and opposite to that in the other  sub-domain. 
 Let us consider a more specific case where the net force exerted by the particle 
 through one sub-domain is ${\bf F}_a$ , given by
\be
	{\bf F}_a = F_{ax} \hat{e}_x + F_{ay} \hat{e}_y.
\ee
Since the particle can only interact with the fluid through the boundary links, it 
can exert force only along those directions.   This force along the boundary links should be chosen such that their sum  is equal to  ${\bf F}_a$.  Obviously, there are many 
ways of distributing this force.  We chose a scheme where the  
 force ${\bf f}_i$ along  each of these links $i$ 
  have components given by 

 \be
 f_{i \alpha}=\frac{c_{i\alpha}}{\sum c_{i\alpha}}F_{a\alpha}. \label{link-force}
 \ee
 where the sum is over all boundary links in the sub-domain  that has a positive component along ${\bf F}_a$.
 
\begin{figure}[!ht]
	\centering
	\includegraphics[scale=0.5,angle=0]{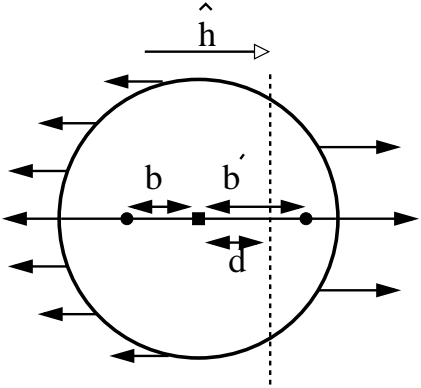}
	\caption{Schematic representation of  the force distribution on the particle.
 The dashed line is the plane which divide the particle  into two. 
The sum of the forces in each part is equal and opposite.  The filled rectangle is
the geometric center and the filled circles are the force centers in each half.}
	\label{schematic-part}
	\end{figure}

   The resulting force distribution is shown, schematically in figure~\ref{schematic-part}.  
These  are the elementary force that will enter into the LB method.

The force acting on each boundary node is then
\be 
{\bf F}_b^i = f_{ix}  {\hat e}_{x}+f_{iy} {\hat e}_y
\ee

This force will result in an additional contribution to the boundary node 
velocity which  can be obtained by inverting eqn~\ref{eqn:fb}.

\be
 {\bf u}_b^a=-\frac{1}{2~a_1^1\rho}\left [ n_i({\bf r},t+1)
             -n_{i^\prime}({\bf r}+{\bf c}_i,t+1)
             -\frac{{\bf F}_b^i\cdot{\bf c}_i}{2c_i^2} \right] {\bf c}_i
\ee

\begin{figure}[!ht]
	\centering
	\includegraphics[scale=0.5,angle=0]{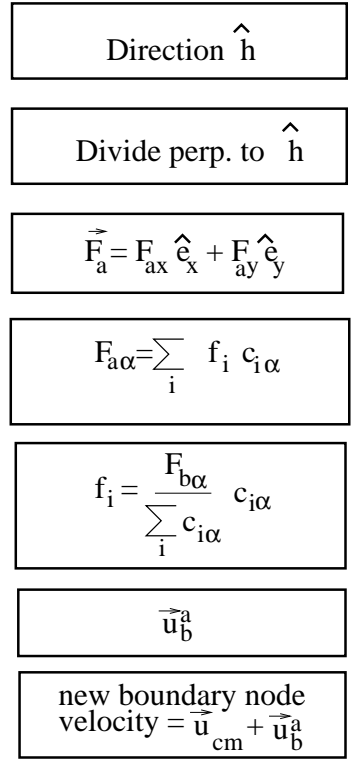}
	\caption{Summary of the procedure for distributing forces on the particle}
	\label{flow}
	\end{figure}

The boundary node velocity that will enter into the calculation of force in eqn
~\ref{force} and torque in eqn~\ref{torque} is now the sum of active 
and passive contributions, ${\bf u}_b+{\bf u}_b^a$. The procedure is summarized in 
figure~\ref{flow}

We restrict ourselves to the zero temperature case,  wherein the fluid 
velocity field is generated solely by the force exerted by the 
active particles on the fluid.  The effect of the parameter $d$, the distance of 
the partitioning line from the geometric center of the particle, is to change the 
number of boundary nodes along which the internal forces are acting. When $d\ne0$ the number of 
terms in the sum in eqn~\ref{link-force} will be different for the two parts of the particle. 
 We will now discuss the implication of this asymmetry in force distribution.

\subsection{Movers and Shakers}

As we discussed before since the forces exerted by the colloid are internal, the 
net force should be zero. This means that the forces exerted by  the boundary nodes
on the fluid, to lowest order,  can only be a force dipole~\cite{brennen-77}.  However, 
the net force exerted by the fluid on the particle depends on the symmetry  of this force 
distribution.

In our simulations, in each part of the particle,  we can define the location of a force center as
	\be
	 {\bf R}_{fc}=\frac{\sum_i {\bf F}^b_{i}\cdot({\bf r}_i{\bf F}^b_{i})}{\sum_i F^2_i}
	\ee
  	where the summation is over all the boundary nodes in that segment.  
If $b$ and $b'$ are the distances to this force center from the geometric center 
of the particle in each half (see figure~\ref{schematic-part})),  then $w=|b-b'|$ is a measure of the
 asymmetry in the force distribution.  The quantities $w$ and $d$ are then linearly related.

When $w$ is non zero,  the velocity field  in the fluid  in the front and back segments of the particle are 
different. This asymmetry in the velocity field results in a net force on the particle which generates its motion. We 
call this a "mover". For single particles  in a symmetric environment this asymmetry in the velocity field 
vanishes  when  $w$ is zero. However the particle still generates a non-zero velocity field in the fluid, 
in other words it behaves as a "shaker".  Stated differently,  if the division of the particle is 
about the geometric center ($d=0$), the force centers in each part are at
 equal distances from the center, and  one obtains an apolar particle, 
otherwise it is polar~\cite{madan-sriram-04}.  Both polar (mover) and apolar (shaker) 
particles disturb the fluid around it.  The apolar particle by symmetry cannot move and will only generate 
a symmetric flow around it (see figure~\ref{shaker-profile}).  But the polar particle will get displaced 
as a result of the applied force distribution, since it can induce an asymmetric flow field around it 
(see figure~\ref{mover-profile}).

\section{ Results - Single Shaker }

\begin{figure}[!ht]
	\centering
	\includegraphics[scale=0.3]{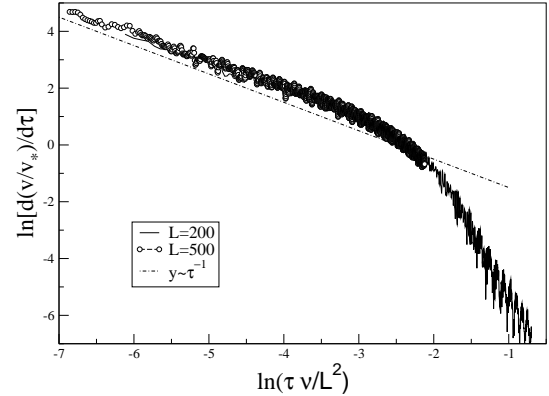}
	\caption{Time derivative of the average speed of the fluid as a function of scaled time,
in the presence of a single shaker. $v_*$ is the average speed at steady state.
Parameters used are $R=5$, $M=1000$ .  All the system sizes simulated fall on the same curve, showing a relaxation with two algebraic decay.}
  \label{time-steady}
	\end{figure}

We will first look at the case of symmetric force distribution on a particle placed at the center 
of the cell \footnote{We work with two values,$M=1000$ and $10000$ for the mass of the particle,  which corresponds to particle densities, $\rho_p=  12 \rho_f$ and $\rho_p=  120 \rho_f$ , respectively. The simulations are stable for both these values of the mass. However for values of $M<1000$ the simulations are unstable.}.  In this case the center of mass of the active particle does not show any 
motion along any preferential direction. As a result of the  force applied by it,  the particle 
will generate a steady velocity profile around it.  In order to sample the statistical properties of the velocity field, we randomly change the  direction  along which force is applied at regular intervals, 
we call this the switching period, and the time interval between them as the switching time. 
We have found that  the maximum fluid disturbance  is achieved if we switch the direction in the time 
scale in which the fluid flow adapts to the perturbation imposed by the particle,  which scales as 
 $L^2/\nu$.   We obtain this time scale as the time taken for the average fluid velocity, to reach a steady value. In figure~\ref{time-steady} we plot the time derivative of the normalized average speed of the fluid particles as a function of time for different values of the system size. As can be seen from the figure, good data collapse is obtained if we scale the time axis by $L^2$. For the rest of this paper we measure time  in units of $L^2/\nu$ . 

\begin{figure}[!ht]
	\centering
	\includegraphics[scale=0.52,angle=0]{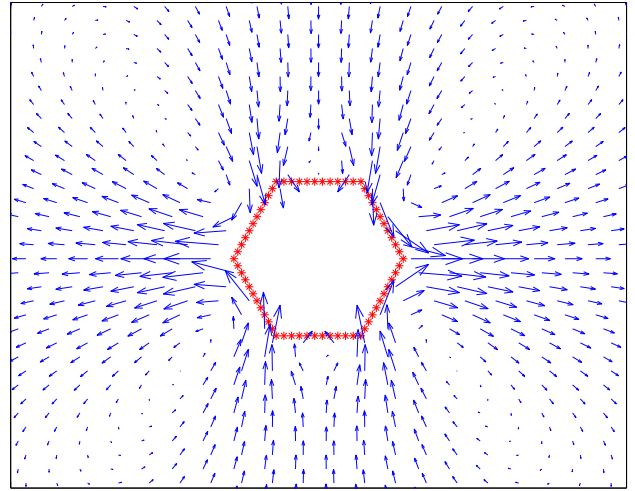}
	\caption{The fluid velocity profile around  a shaker. $R$=5, $L$=100, and $M$=1000}
	\label{shaker-profile}
	\end{figure}

\begin{figure}[!ht]
	\centering
	\includegraphics[scale=0.45,angle=0]{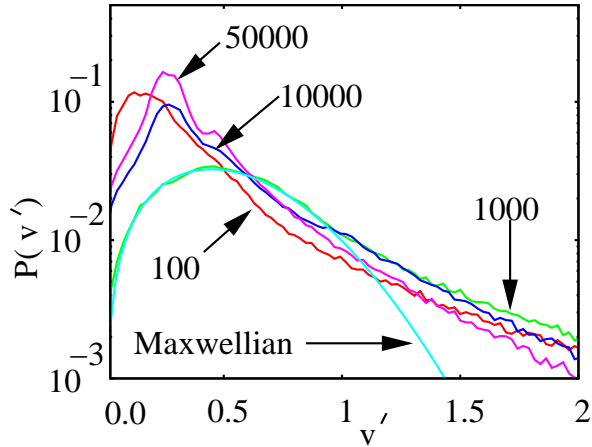}
	\caption{Normalized fluid velocity distribution  $P(v^{\prime})$ vs $v^{\prime}$ for different 
	switching times in a semi-log plot for a shaker.  $v^{\prime}$is the velocity  scaled by 
	the average speed of the  fluid velocity. Other parameters are   $R$=5, $L$=100 and  $M$=10,000}
        \label{shaker-pofv}
	\end{figure}

The velocity profile around the particle at this steady state condition is 
shown in figure~\ref{shaker-profile}. The force applied by the particle on the fluid creates four vortices that are symmetrically placed. As a result, there is no net  force acting  on the particle and it does not get displaced.
In figure ~\ref{shaker-pofv} we show the fluid velocity distribution for different switching rates. It is clear
 from the figure that there is a dependence on the switching rate. Maximum disturbance in the fluid is created 
when the switching rate is roughly $1/(10 L^2/\nu)$,  which corresponds to a 
switching time of the order  the  time  needed to reach the steady state . The  velocity distribution in the steady departs clearly from a Maxwellian. This is expected given the fact that the system is  clearly out of equilibrium. In figure~\ref{shaker-pofv} we compare a Maxwellian  with  the distribution obtained from the simulations  to show that except for very slow switching rates we have significant deviation from it at large velocities. 

\section{Results - Single Mover }

\begin{figure}[!ht]
	\centering
	\includegraphics[scale=0.52,angle=0]{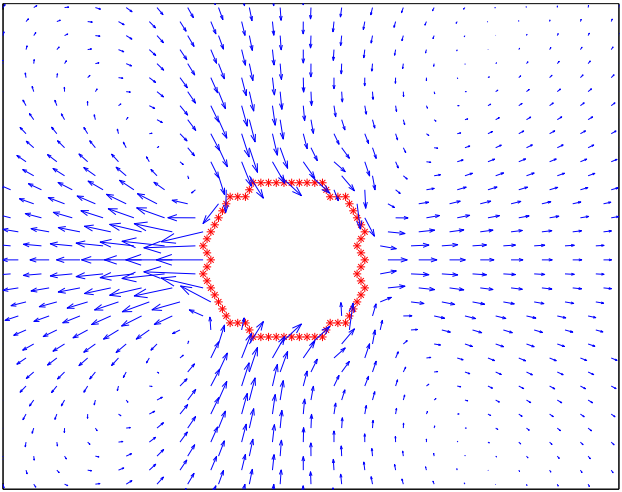}
	\caption{ Asymmetric velocity profile of the fluid around a mover at steady state. $R$=5, $L$=100 and $M$=1,000 }
	\label{mover-profile}
	\end{figure}

        When the force centers   are not symmetrically 
placed about the particle's geometric center ( $w\ne0$)
the interaction of the particle  with the fluid results in a  net force 
 on the particle.   The particle now has a direction set by the vector $\vec{d}$, along which it starts to move. 
 Starting from rest, the particle and the fluid  reach a steady state velocity on a time scale $L^2/\nu$, as was the case for the shaker.  However the fluid velocity profile, at this steady 
  state,  is quite different from that of the shaker. As can be seen in figure~\ref{mover-profile}
  the vortices in the  front and back of the particle are now asymmetric, clearly indicating the direction 
  of motion of the particle. 

\begin{figure}[!ht]
	\centering
	\includegraphics[scale=0.45,angle=0]{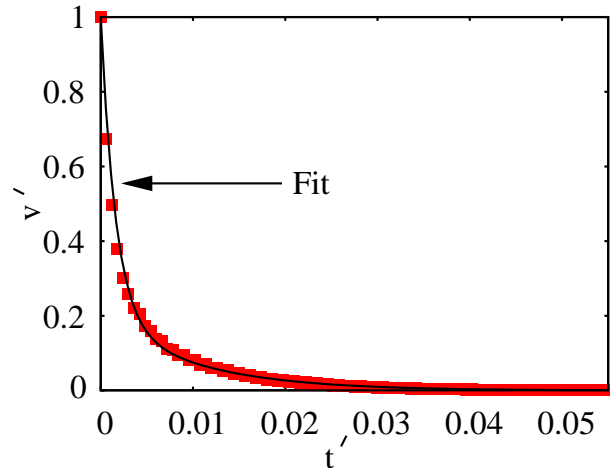}
	\caption{ The velocity of the colloidal particle  as a function of time after the driving 
	is switched off. The continuos line is a fit to the function of the form 
	$v_1 e^{-t^\prime/\tau_1}+v_2 e^{-t^\prime/\tau_2}$ with  $\tau_1=0.0016$ , $\tau_2=0.01$ , $v_1=0.8$ and $v_2=0.2 $. The data is for a particle of radius $R=7$, 
	mass $M=1000$  . The size of
	the simulation box is $L=100$.   The  velocity has been scaled by  the particle velocity 
	at the time when the activity was switched off i.e.at $t^\prime=0$  }
	\label{dissip}
	\end{figure}

We have also analyzed the decay of the particle's velocity if its activity is switched off. For the systems analyzed, such a decay is consistent with  a double exponential.
This is depicted in the figure~\ref{dissip}. 
The time scales $\tau_1$ and $\tau_2$ originate from different dissipation mechanisms. The 
first time scale comes from the friction between the particle and the surrounding fluid, and is the 
result of particle's inertia.  In a short time 
after  activity is switched off, the velocity profile changes to 
one with no slip at the boundary. Further dissipation is due to the  drag resulting from the distortions in the fluid created by the particle motion.  These distortions travel
through the fluid and are eventually dissipated. In two dimensions, the resistance offered to  the particle motion due to  this Stokes' drag,  has a weak dependence on both the system size and the particle size~\cite{punkkinen}.

\begin{figure}[!ht]
	\centering
	\includegraphics[scale=0.45,angle=0]{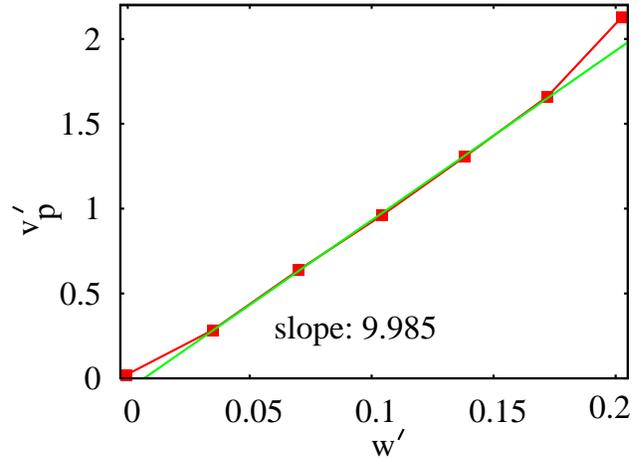}
	\caption{Speed of the particle for different values of $w$. Other parameters are $R$=5, $L$=100 and $M$=1000.  Here $\langle v_p^{\prime} \rangle$ is the average particle particle velocity in units of  $L/\nu$}
	\label{dipole-vel}
	\end{figure}

The velocity of the particle depends on two factors, the strength of the force in 
each half ($f$) and the asymmetry in the force center positions with respect to the 
geometric center $(w)$.  The dependence of the particle velocity on $w$ 
is shown in figure~\ref{dipole-vel}

	\begin{figure}[!ht]
	\centering
	\includegraphics[scale=0.45,angle=0]{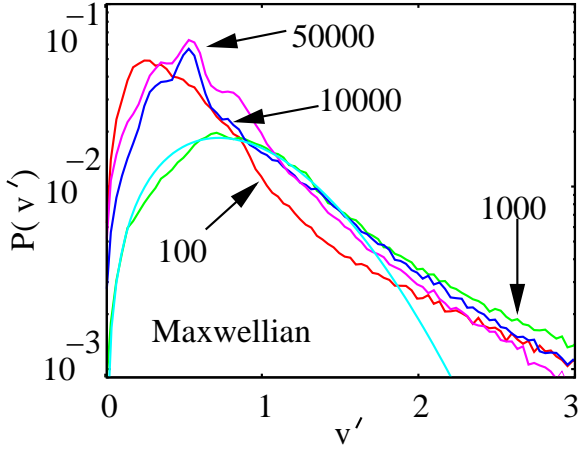}
	\caption{Normalized velocity distribution $P(v^{\prime})$ vs $v^{\prime}$ of the fluid nodes for 
	different switching times (shown with arrow marks) for a mover.  Velocity, $v^{\prime}$,  is the 
	velocity scaled by the average particle velocity. Other parameters are   $R$=5, $L$=100 and  $M$=10,000 }
	\label{mover-vel-prof}
	\end{figure}
       
     As happens with  the shaker, the activity of the particle results in a non-Maxwellian distribution 
      for the fluid node velocities. This is shown in figure~\ref{mover-vel-prof}. This distribution also
        has a dependence on the switching time of the particle, with the maximum velocity obtained when the switching time is comparable the the diffusion time set by $L^2/\nu$. The curves are analogous to those obtained in fig.~\ref{shaker-pofv}, except for the fact that the tail is longer here.	
        
	\begin{figure}[!ht]
	\centering
	\includegraphics[scale=0.45,angle=0]{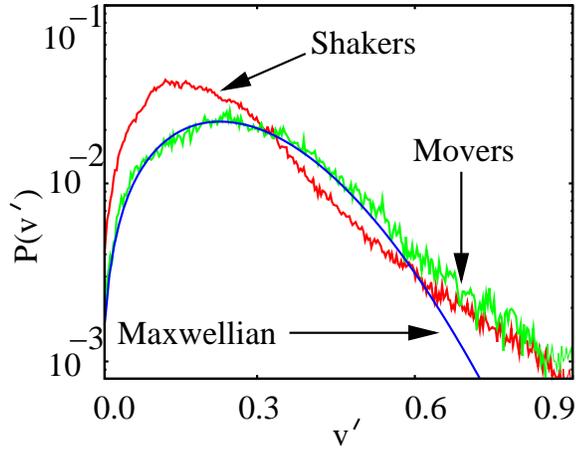}
	\caption{$P(v^\prime)$ vs $v^\prime$ for five movers and shakers in a semi-log plot. 
	Other parameters are the same as in ~\ref{mover-vel-prof}. }
	\label{pofv-nop5}
	\end{figure}

	\begin{figure}[!ht]
	\centering
	\includegraphics[scale=0.45,angle=0]{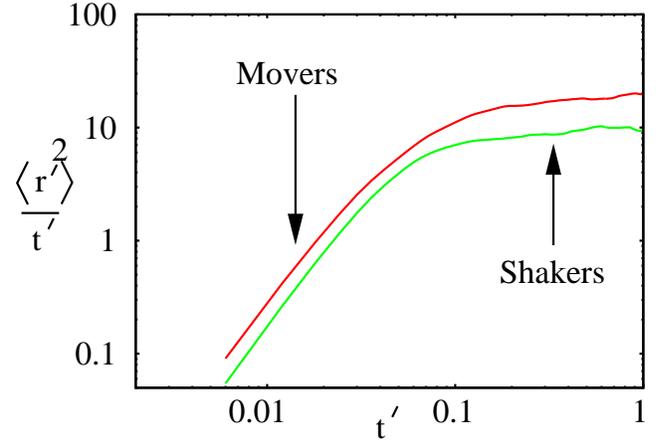}
	\caption{$\frac{\langle r^2\rangle}{t^\prime}$ vs $t^\prime$ with five active particles and one 
        passive particle.  The position of the passive particle is tracked as a function of time
	to obtain the data. $R$=5, $L$=100 and $M$=10,000 } 
	\label{diffusion}
	\end{figure}

\subsection{\bf Many Particles}
     We will now look at the effect that  a suspension of active particles  have 
     on the fluid. We 
restrict ourselves to the dilute limit where direct interactions have a negligible contribution.
 
     The fluid velocity distribution  with five active 
particles in a $100 \times 100 $  lattice is presented in figure
\ref{pofv-nop5} for both movers and shakers .  The  
non-Maxwellian  nature of the distribution is highlighted here by comparing 
the $p(v)$ for large $v$ with that obtained from the best fit Maxwell distribution.
 
 In order to further check the effect of shakers and movers on the 
fluid we compute the diffusion coefficient of  a passive bead, having the 
same radius as the active ones, suspended in the fluid along with several 
other active particles. The mean
square displacement of this particle as a function of time is shown in 
figure~\ref{diffusion}.
During early times, the 
$\langle r^2 \rangle$ shows ballistic behavior, and at later times a crossover regime  leads eventually to a purely diffusive regime.

 The above two measurement  indicate that the movers keep the system at a 
 higher temperature than the shakers. 
 
\section{Conclusion}

We have presented a model for the numerical simulation of self-propelled
colloidal suspensions based on the Lattice-Boltzmann method for colloids.
By devising a way of adding internal forces to the boundary interaction
between the fluid and the solid particle so as to satisfy the net zero force
condition, we are able to model two types of particles - movers and shakers.
A symmetric force distribution results in a shaker particle, which disturbs the
fluid around it by producing a velocity field but without producing any net displacement.  A mover is
a result of asymmetric force distribution. These particles can move by
using the asymmetry in the velocity field around it. We show that the
velocity of the mover is directly proportional to the asymmetry in
the position of the force distribution.

 The statistical properties of the velocity field generated by the
activity is shown to deviate from that of equilibrium fluids, with the velocity distribution having a generalized exponential  exponential tail $p(v)~\exp(-(v/v_0)^{\alpha}$.  It is shown that the momentum diffusion and the different dissipative mechanisms 
are consistent with what is well known in two dimensional fluids. 

The model could be used to study the emergence of collective behavior
in self propelled organisms as well as the rheological properties of
the suspension.  These aspects are currently under investigation.

\bibliographystyle{prsty}
\bibliography{active}
\end{document}